УДК 004.891

# ЭКСПЕРИМЕНТАЛЬНАЯ ПРОВЕРКА МОДЕЛИ ОЦЕНКИ ИННОВАЦИОННОСТИ ОБЪЕКТА

## В.К. Иванов

*Тверской государственный технический университет, 170026, г. Тверь, наб. Аф. Никитина, д. 22, mtivk@mail.ru*

**Аннотация:** В статье рассматривается подход к количественной оценке инновационности продуктов и технологий. Результаты такой оценки могут быть использованы при создании хранилища данных для описаний объектов со значительным инновационным потенциалом. Модель расчета индекса инноваций основана на понятиях новизны, актуальности и имплементируемости объекта. Даны формальные определения этих показателей и описана методика их расчета. Используются нечеткие методы для обработки (неполной) информации из многочисленных источников и для получения вероятностных оценок инноваций. Представлены экспериментальные данные проверки модели, в том числе расчеты локальных критериев и глобального аддитивного оценочного критерия. Установлены цикличность динамических изменений показателей, их взаимозависимость, некоторые общие особенности продвижения продуктов. Полученные экспериментальные данные согласуются с экспертными оценками исследуемых продуктов. Анализ локальных критериев, использованных в исследовании, дает основание утверждать о правильном использовании аддитивной n-мерной функции полезности. Адекватность предположений и формальных выражений, которые используются в вычислительных алгоритмах отбора информации для размещения в хранилище данных, подтверждается.

**Ключевые слова:** хранилище данных, инновационность, аддитивный критерий, функция полезности, поисковый запрос.

## ВВЕДЕНИЕ

В проекте "Организация и поддержка хранилища данных на основе интеллектуализации поискового агента и эволюционной модели отбора целевой информации" [1] предлагается технология создания хранилища описаний объектов (продуктов и технологий), обладающих значимым инновационным потенциалом. Инновационный потенциал количественно оценивается индексом инновационности, модель вычисления которого основана на показателях новизны, востребованности и имплементируемости объекта. Указанные показатели вычисляются для лингвистической модели объекта, создаваемой с помощью архетипов.

Архетипы объекта (АО) – это концепции предметной области для рассматриваемого объекта. АО реализуются термами, определяющими ключевые свойства объекта, и группируются в три дескриптивных класса: структура объекта, условия применения и результаты функционирования. Область определения АО определяется термом-маркером. Также могут существовать дополнительные локальные ограничения: умолчания, синонимы термов, веса термов, предельное количество запросов, количество термов в запросе и т.п. Классы АО, маркер и локальные ограничения образуют лингвистическая модель объекта. Лингвистическая модель используется как поисковый паттерн для генерации набора запросов на поиск информации о потенциально инновационном объекте.

Поисковые запросы - логические выражения, где множества операндов есть различные комбинации термов-архетипов и маркера. Релевантные документы, найденные после исполнения всех сгенерированных запросов, используются для вычисления показателей инновационности объекта.

Цель статьи – подтвердить основные предположения, используемые при разработке модели для вычисления аддитивного оценочного критерия инновационности объектов, представив данные экспериментальной проверки этой модели.

## РАБОТЫ ПО ТЕМАТИКЕ ИССЛЕДОВАНИЯ

Обсуждаемые в настоящей статье глобальный и частные критерии применяются в разрабатываемой технологии создания хранилища описаний инновационных объектов. Анализ источников [2], [3], посвященных различным аспектам инновационного развития, показывает, что авторы в понятие "инновация" всегда включают такие коннотаты, как: новый, научный, повышающий эффективность, приносящий прибыль. Исходя из этого, далее в статье формализуется понятие инновационность.

Теоретической основой формирования многокритериальных скалярных оценок, включая глобальный аддитивный критерий, является аксиоматическая теория полезности [4]. Одним из важных положений теории полезности является доказанное утверждение, что из аддитивности оценочной функции следует взаимная независимость факторов, влияющих на нее [5]. Анализ используемых в исследовании частных критериев дает основание утверждать о корректном использовании аддитивной n-мерной функции полезности.

Так как ожидается очевидная неполнота и неточность информации об инновационном потенциале объектов, полученной из различных источников, использованы подход и методы теории свидетельств [6], [7].

Эта статья является продолжением предыдущих работ автора и его коллег о различных аспектах технологии создания хранилища описаний инновационных объектов [8], [9] и др.).

## МОДЕЛЬ ВЫЧИСЛЕНИЯ ИНДЕКСА ИННОВАЦИОННОСТИ ОБЪЕКТА

### Основные понятия

Ниже сформулированы понятия новизны, востребованности и имплементируемости, как составных частей критерия инновационности искомого объекта. Количественная оценка этих показателей основывается на предположении об адекватности отображения жизненного цикла продуктов в виде информационных объектов, размещенных в различных хранилищах данных.

*Новизна* определяет значительные улучшения, новый способ использования или предоставления объекта. Предполагается, что для новых объектов количество найденной информации будет меньше, чем для давно существующих.

*Востребованность* (или спрос) – это осознанная потенциальным потребителем необходимость в этом объекте. Оценка востребованности основывается на значении частоты запросов пользователей к информации об объекте, находящейся в хранилищах данных. Примеры: частота выполнения запросов, похожих на запросы из поискового паттерна; показатель цитирования материалов об объекте; количество продаж объекта.

*Имплементируемость* определяет технологическую обоснованность, физическую осуществимость и способность объекта быть интегрированным в какую-либо систему для получения желаемых эффектов. Оценка имплементируемости основывается на значении среднего периода восстановления уровня новизны и/или востребованности объекта после их потери со временем. Чем быстрее это происходит за счёт новых технологий, конструкций, улучшенных функциональных и потребительских характеристик, тем выше имплементируемость.

**Базовые вычисления**

Индикатор технологической новизны $Nov$ вычисляется следующим образом:

$$Nov = 1 - 1/N \sum_{k=1}^{N} f_n^{01}(R_k,...), \qquad (1)$$

где $N$ – общее количество выполненных запросов; $R_k$ – число документов, найденных в результате выполнения $k$-го запроса; $f_n^{01}(R_k,...)$ – вариативная функция, нормирующая значение $R_k$ на диапазон [0;1].

Индикатор востребованности $Dem$ вычисляется следующим образом:

$$Dem = 1/S \sum_{k=1}^{S} f_n^{01}(F_k,...), \qquad (2)$$

где $S$ - общее количество выполненных запросов; $F_k$ – частота выполнения $k$-го запроса; $F_{01}(F_k,...)$ – вариативная функция, нормирующая значение $F_k$ на диапазон [0;1].

Индикатор имплементируемости $Imp$ вычисляется следующим образом:

$$Imp = 1 - 1/2(\Delta t_N(Nov(t)) + \Delta t_D(Dem(t))), \qquad (3)$$

где $Nov(t)$ – функция, показывающая зависимость $Nov$ от времени на временном интервале $[t_0; t_m]$; $Dem(t)$ – функция, показывающая зависимость $Dem$ от времени на том же $[t_0; t_m]$; $\Delta t_N$ и $\Delta t_D$ – средние расстояния между двумя последовательными точками временного ряда $t_i, t_{i+1} \in [t_0; t_m]$ локальных максимумов функций $Nov(t)$ и $Dem(t)$. При этом $Nov$, $Dem$ и $Imp$ рассчитываются для точки $t_{m+1}$.

Индекс инновационности $Ix$ имеет вид аддитивного критерия:

$$Ix = w_{Nov}Nov + w_{Dem}Dem + w_{Imp}Imp, \qquad (4)$$

где $w_{Nov}$, $w_{Dem}$, $w_{Imp}$ - весовые коэффициенты для $Nov$, $Dem$, и $Imp$ соответственно и $w_{Nov} + w_{Dem} + w_{Imp} = 1$.

**Вычисления при неполной и неточной информации об объектах**

В этом случае в проекте вводятся нечеткие показатели вероятности того, что объект обладает новизной, востребованностью и имплементируемостью. Для вычисления указанных вероятностей применяется теория свидетельств Демпстера-Шафера [6], [7]. Определяются базовые вероятности $m$ попадания результатов измерения $Nov$, $Dem$, и $Imp$ в $k$-й интервал значений $A_k$; результаты из различных источников комбинируются с помощью разработанного специализированного алгоритма групповой обработки результатов измерений. Комбинирование выполняется рекурсивно по парам источников: из двух источников свидетельств образуется один условный источник, свидетельства которого комбинируются с очередным фактическим источником.

Рассчитываются функция доверия $Bel(A) = \sum_{A_k : A_k \subseteq A} m(A_k)$ и функция правдоподобия $Pl(A) = \sum_{A_k : A_k \cap A} m(A_k)$, которые определяют верхнюю и нижнюю границу вероятности обладания объектом свойства, заданного соответствующим фактором. Тогда выражение (4) приобретает вид мультипликативного оценочного критерия:

$$Ix = [Bel_{Nov}(A), Pl_{Nov}(A)]^{w_{Nov}} * [Bel_{Dem}(A), Pl_{Dem}(A)]^{w_{Dem}} \\ * [Bel_{Imp}(A), Pl_{Imp}(A)]^{w_{Imp}}, \quad (5)$$

которое сводится логарифмированием $Ix$ к аддитивному критерию:

$$\ln Ix = w_{Nov} * \ln([Bel_{Nov}(A), Pl_{Nov}(A)]) + w_{Dem} * \\ \ln([Bel_{Dem}(A), Pl_{Dem}(A)]) + w_{Imp} * \ln([Bel_{Imp}(A), Pl_{Imp}(A)]) \quad (6)$$

Так как $\ln()$ возрастающая функция, рассуждения, касающиеся $Ix$, справедливы для $\ln(Ix)$.

## МЕТОДИКА И РЕЗУЛЬТАТЫ ЭКСПЕРИМЕНТАЛЬНЫХ ИССЛЕДОВАНИЙ

### Методика проведения экспериментов

Цель проведенных экспериментов – проверить модель вычислений локальных и глобального оценочных критериев и тем самым подтвердить адекватность формальных выражений (1), (2) и (3), которые используются в вычислительных алгоритмах отбора информации для размещения в хранилище данных.

В качестве объектов с очевидным инновационным потенциалом были выбраны смартфоны популярных моделей известных производителей. После этого экспертами были сформированы две лингвистических модели: *iPhone X* и *Samsung Galaxy S*. Эти модели были использованы для генерации поисковых запросов. Пример поисковых запросов (маркер - «смартфон iPhone»; АО: «камера», «экран», », «музыка», «производительность», «аккумулятор»; ограничение – 3 терма в запросе):

*смартфон AND iPhone AND камера*

*смартфон AND iPhone AND экран*

*смартфон AND iPhone AND производительность*

*смартфон AND iPhone AND музыка*

*смартфон AND iPhone AND аккумулятор*

В качестве источников данных об объектах использовались следующие хранилища данных: Google Scholar (https://scholar.google.ru), AliExpress (https://aliexpress.com), ACM Digital Library (https://dlnext.acm.org), IEEE Explore Digital Library (https://ieeexplore.ieee.org).

Измеряемые показатели: *Nov* , *Dem* (усреднялись среднеарифметическим и медианным значениями, нормализовались на диапазон [0;1] функциями для линейной и экспоненциальной нормализации. Вычисляемые показатели: *Imp* , *Ix* .

**Результаты экспериментов**

На рис. 1 представлен график изменения индикатора новизны исследуемых объектов за 10-летний период. На этом и последующих графиках аппроксимирующие кривые показаны пунктиром.

На рис. 2 представлено сравнение графиков изменения индикатора новизны объектов в зависимости от вида нормирующей функции.

На рис. 3 и 4 представлены графики изменения индикатора востребованности исследуемых объектов за 10-летний период (для случаев, где $F_k$ - медиана и среднее арифметическое).

На рис. 5 представлена сравнительная диаграмма индикаторов новизны, востребованности и имплементируемости исследуемых объектов, построенная по данным из двух источников.

На рис. 6 представлена сравнительная диаграмма индикатора востребованности исследуемых объектов в сравнении с экспертной оценкой рейтинга смартфонов (https://rskrf.ru/ratings/tekhnika-i-elektronika/electronic/smartfony-) «Российской системы качества».

В ходе проведения экспериментов было получен достаточно данных для первичного анализа динамики изменений показателей инновационности объектов и сравнения их с экспертными оценками. Рис. 1-6 иллюстрируют часть этих данных.

## ОБСУЖДЕНИЕ РЕЗУЛЬТАТОВ

Очевидна цикличность изменений анализируемых показателей (рис. 1-4), которая соответствует предположениям о наличии инновационных циклов в конкретной области применения.

1. Для рассмотренной области применения (смартфоны) установленная цикличность появления инновационных продуктов – 5-6 лет (рис. 1-4). Эта периодичность требует дополнительного анализа.
2. Анализ соотношений показателей новизны о востребованности объектов выявляет определенные закономерности. Падение востребованности вызывает рост новизны объекта, а рост востребованности объекта приводит к падению его новизны. Подтверждение см. на рис. 1-3.

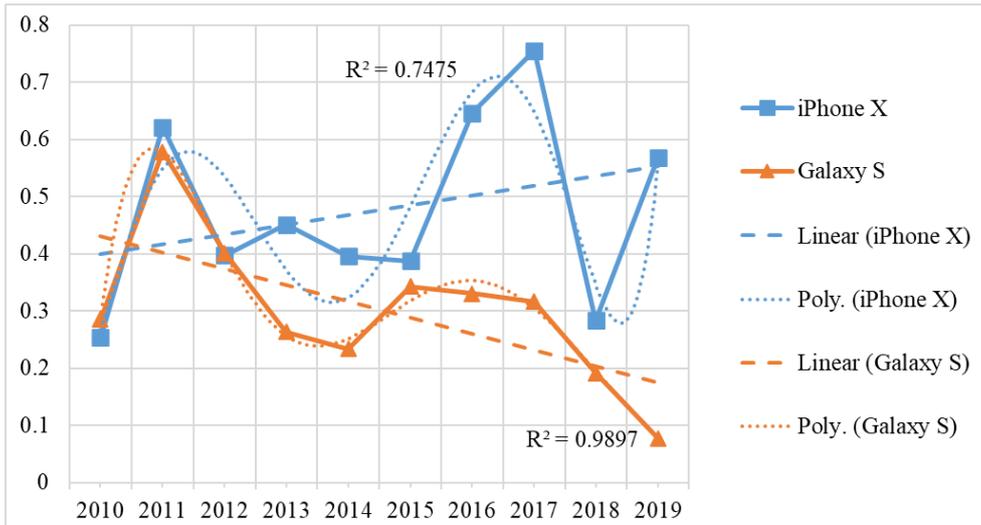

Рис.1. Динамика изменения новизны объектов

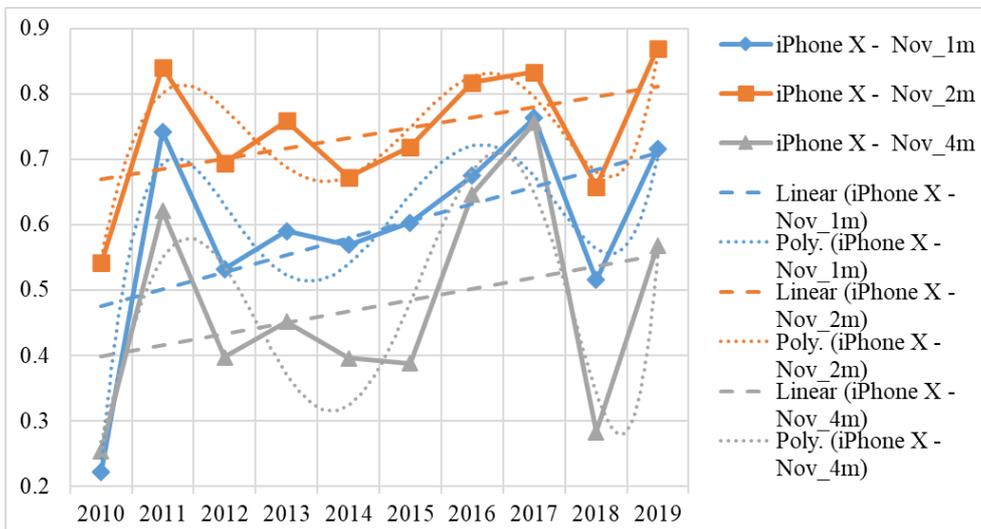

Рис.2. Динамика изменения новизны объектов для различных нормирующих функций

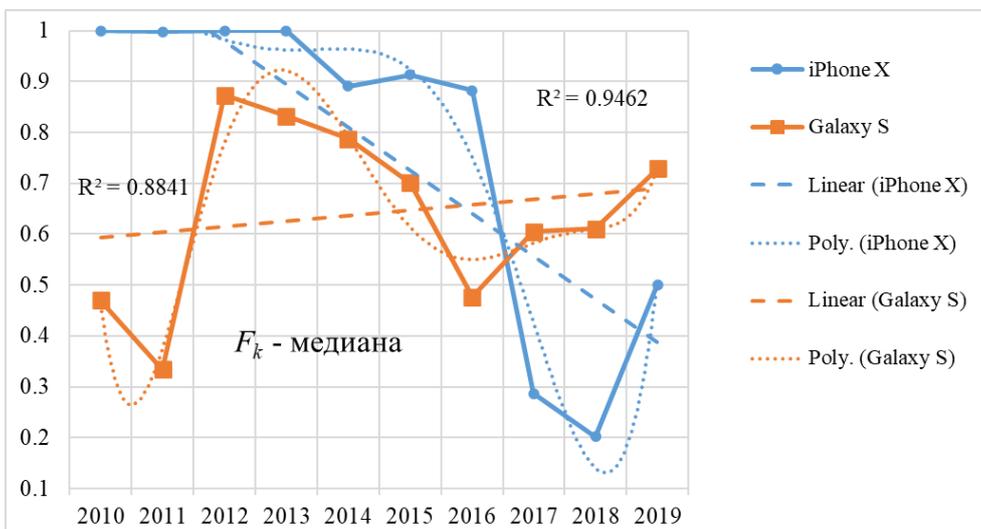

Рис.3. Динамика изменения востребованности объектов

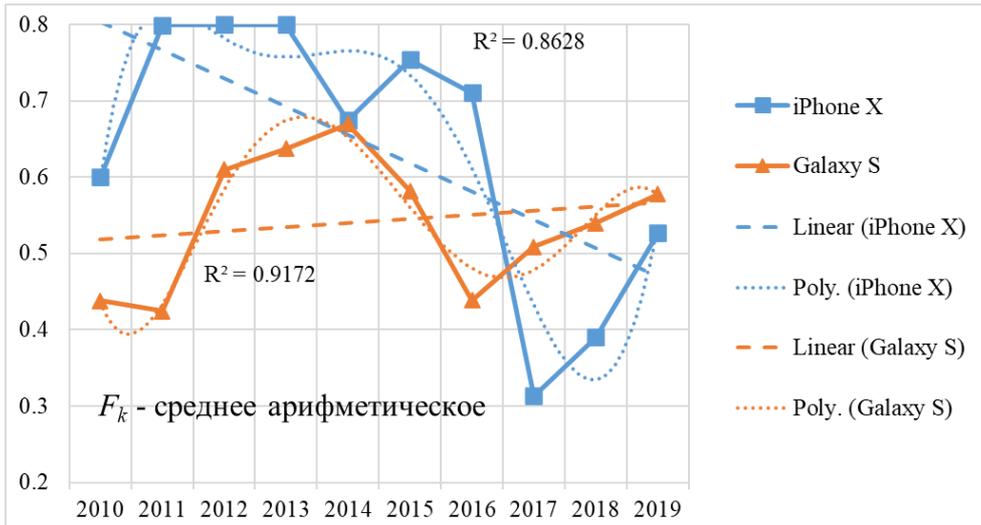

Рис.4. Динамика изменения востребованности объектов

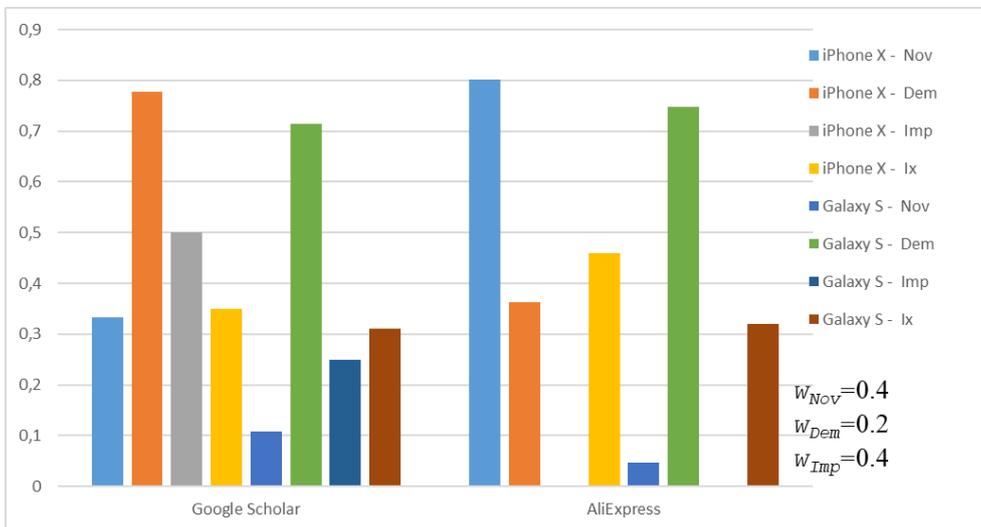

Рис.5. Индикаторы инновационности исследуемых объектов

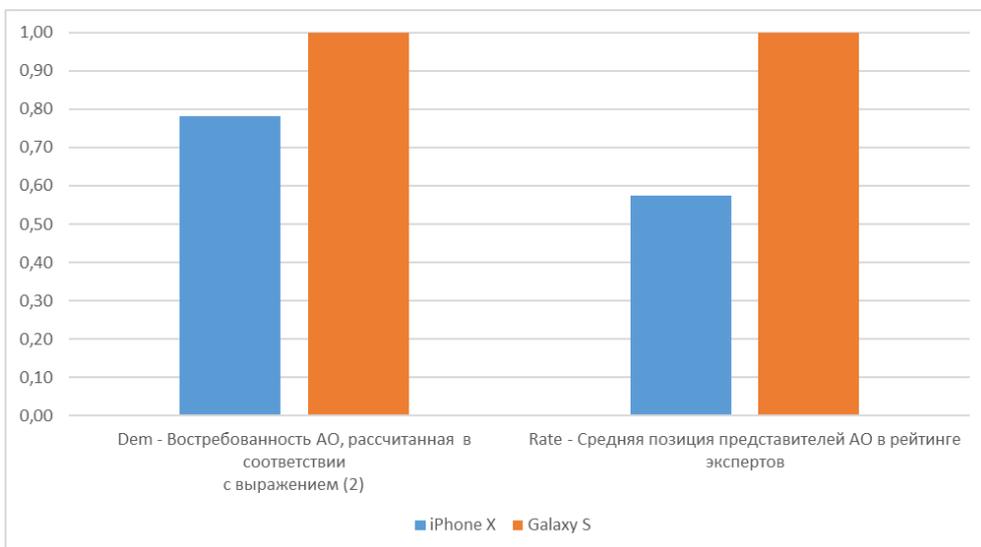

Рис.6. Сравнение вычисленного индикатора востребованности объекта с экспертной оценкой

3. Полученные данные позволяют оценить некоторые особенности производства и продвижения продуктов. Так, восходящий тренд новизны и нисходящий тренд востребованности продуктов *iPhone X* показывают высокий производственный потенциал производителя и наличие резервов. в маркетинге. Для продуктов *Samsung Galaxy S* идентифицируется наличие производственных резервов(нисходящий тренд новизны) и стабильный уровень маркетинга (флэт востребованности продуктов). См. рис. 1, 3 и 4.
4. Особенностью аддитивного критерия является возможность взаимной компенсации значений локальных критериев при вычислении глобального оценочного критерия. Эта ситуация показана на рис. 5. Здесь значения показателей *Nov* и *Dem* для продуктов *iPhone X* существенно отличаются для различных источников исходных данных. Кроме того, соотношение между значениями как *Nov* так и *Dem* обратны для этих источников. Однако значения глобального критерия *Ix* для обоих источников сопоставимы.
5. Аппроксимация полученных значений показателей инновационности позволяют оценить (и в определённом смысле прогнозировать) их динамику.
6. Чувствительность модели вычислений к виду нормирующей функции очевидна – существенно различаются значения *Nov* (см. рис. 2). Отметим некоторые особенности. Линейная аппроксимация дает практически одинаковый угловой коэффициент для всех трех вариантов, но различные величины сдвига. Полиномиальная аппроксимация показывает одинаковую периодичность *Nov* для всех трех вариантов, что говорит об отсутствии влияния нормирующей функции на вычисление значения *Imp*. Более точная оценка условий применения методов нормирования требует дополнительных исследований.
7. Полученные экспериментальные данные хорошо согласуются с экспертными оценками исследуемых объектов (рис. 6). Показано соотношение показателей востребованности разных объектов, вычисленных в соответствии с (2). Аналогичное соотношение наблюдается при анализе результатов экспертизы. Результаты вычислений и экспертизы востребованности объектов вполне сравнимы, так как методика экспертных оценок смартфонов предусматривала выбор параметров оценки и их значимости, исходя из предпочтений пользователей.
8. Теоретическое обоснование типа оценочного критерия базируется на положениях теории полезности и теории свидетельств. Их достаточная апробированность позволяет надеяться на достаточную адекватность разработанных моделей, учитывая одновременное использование эвристических подходов.

## ЗАКЛЮЧЕНИЕ

Представленная в статье модель используется при разработке вычислительных алгоритмов для отбора информации об объектах с инновационным потенциалом. Экспериментальные данные позволили усовершенствовать методику расчёта показателей инновационности объектов, включая совместную обработку нечётких данных, полученных из разных поисковых систем. Результаты исследования подтвердили основные предположения, используемые при разработке модели. В дальнейших исследованиях предполагается расширить экспериментальную базу для получения более статистически значимых результатов.

## БЛАГОДАРНОСТИ

# EXPERIMENTAL CHECK OF MODEL OF OBJECT INNOVATION EVALUATION

## Vladimir K. Ivanov

*Tver State Technical University, Af. Nikitina emb. 22, Tver, Russia, mtivk@mail.ru*

**Abstract.** The article discusses the approach for evaluating the innovation index of the products and technologies. The evaluation results can be used to create a warehouse of the object descriptions with significant innovation potential. The model of innovation index computation is based on the concepts of novelty, relevance, and implementability of the object. Formal definitions of these indicators are given and a methodology for their calculation are described. The fuzzy methods to coprocess (incomplete) data from numerous sources and to obtain probabilistic innovation assessments are used. The experimental data of the model check including the calculations of local criteria and global additive evaluation criterion are presented. The cyclical nature of dynamic changes in indicators, their interdependence was established, some general features of the products promotion were found. The obtained experimental data are consistent with expert estimates of the products under study. The analysis of the local criteria used in the research gives grounds to assert the correct use of the additive n-dimensional utility function. The adequacy of assumptions and formal expressions that are used in computational algorithms for selection information for data warehouse is confirmed.

**Keywords:** Data Warehouse, Innovation, Additive Criterion, Utility Function, Search Query.

# СВЕДЕНИЯ ОБ АВТОРЕ

Об авторе: **Иванов Владимир Константинович** — к. т. н.; доцент; начальник Управления информационных ресурсов и технологий; **ФГБОУ ВО «Тверской государственный технический университет» (ТвГТУ)**; 170026, г. Тверь, наб. Аф. Никитина, д. 22, офис Ц-205; mtivk@mail.ru; +79051272231.

About the author: **Ivanov Vladimir Konstantinovich** - Associate professor; Ph. D. (Engineering) sciences; Director of Center for Research and Educational Digital Resources; **Tver State Technical University (TvSTU)**; Office 205, 22 Afanasy Nikitin Embankment, Tver, 170026, Russia; mtivk@mail.ru; +79051272231.